\documentstyle[preprint,prc,aps]{revtex}

\preprint{IU/NTC\ \ 01--11}

\tightenlines

\begin{document}

\title{
Covariant Effective Field Theory\\
         for Bulk Properties of Nuclei}

\author{Brian D. Serot}

\address{Physics Department and Nuclear Theory Center, 
Indiana University\\
Bloomington, IN\ 47405, USA}

\maketitle

\bigskip
\begin{abstract}
Recent progress in Lorentz-covariant quantum field theories of the
nuclear many-body problem ({\em quantum hadrodynamics}, or QHD) is
discussed.
The importance of modern perspectives in effective field theory
and density functional theory for understanding the successes
of QHD is emphasized.
The inclusion of hadronic electromagnetic structure and of
nonanalytic terms in the energy functional is  also considered.
\end{abstract}

\label{sec:relativistic}
\def\Sch{Schr\"odinger}

\section{Introduction}
\label{sec:relintro}

A central goal of nuclear theory is to describe atomic
nuclei in terms of their constituents.
This goal is especially relevant now, in view of the new accelerators that
will probe nuclei with high energies and high precision using electrons,
hadrons, and heavy ions.
These accelerators are expected to reveal new physics involving the 
properties of hadrons inside nuclei, the nuclear matter phase diagram, 
the role of relativity, and the dynamics of the quantum vacuum.
Moreover, we hope to learn not only about hadron dynamics but also
about the role of explicit quark and gluon degrees of freedom.

Although the \Sch\ equation has been basically successful in describing 
nuclei at low energies,
this framework must be extended if we are to compare calculations
with the data of the future.
A more complete treatment of hadronic systems should include relativistic
nucleon--nucleon interactions, 
dynamical mesons and baryon resonances, modifications
of the hadron structure in the nucleus, and the dynamics of the quantum
vacuum, while respecting the general principles of quantum mechanics, Lorentz
covariance, gauge invariance, cluster decomposition, and microscopic causality.
These physical effects will be relevant regardless of the degrees of freedom
used to describe the system, and they must be studied simultaneously and
consistently to draw definite conclusions about nuclear dynamics at high
temperatures, high densities, and short distances.

In principle, quantum chromodynamics (QCD) should provide such a description,
since QCD is the fundamental theory of the strong interaction.
Nevertheless, QCD predictions at nuclear length scales with the precision of
existing (and anticipated) data are not now available, and this state of
affairs will probably persist for some time.
Even if it becomes possible to use QCD to
describe nuclei directly, this description is likely to be awkward, since 
quarks cluster into hadrons at low energies, and hadrons (not quarks or 
gluons) are the degrees of freedom actually observed in experiments.

Given the present inadequacy
of both the \Sch\ equation and QCD for formulating an
improved description of nuclear dynamics, we must consider alternatives.
Since hadrons are the relevant experimental degrees of freedom, it is
important to see if practical, reliable, and accurate hadronic descriptions
can be developed for the energy, density, and temperature regimes
obtainable with the new experimental facilities.
In any lagrangian approach, one must first decide on the generalized
coordinates, and hadronic degrees of freedom---baryons and mesons---are the
most appropriate for the vast majority of nuclear phenomena.

Lorentz-covariant, meson--baryon effective
field theories (``quantum hadrodynamics'' or QHD) have proven to be
useful for quantitative descriptions of the nuclear many-body 
problem \cite{SW86r,BDS92r,FST97r,SW97r,FSp00r,FSL00r,EvRev00r,SW00r}.
When applied within the framework of modern effective field theory (EFT)
and density functional theory (DFT), they provide a
realistic description of bulk nuclear properties and the spin-orbit
force throughout the Periodic Table \cite{SW97r,FRS98r,FSL00r}.
This success arises from the presence of large scalar and vector mean
fields, which imply that there are large relativistic interaction
effects in nuclei under normal conditions \cite{EvRev00r}.
There is evidence from QCD sum rules that these large fields are
dynamical consequences of the underlying 
chromodynamics \cite{COHEN91r,FURNST92r}.
Moreover, similar relativistic effects are responsible for the efficient
description of spin observables in medium-energy proton--nucleus
scattering using the Relativistic Impulse Approximation, and they
are consistent with the major role played by scalar and vector meson
exchange in modern boson-exchange models of the nucleon--nucleon (NN)
interaction.
All of these features motivate further investigations into the application
of QHD to the nuclear many-body problem.

Since QHD contains strong couplings, one needs a consistent
underlying framework and a systematic expansion procedure to perform
reliable calculations.
One possible framework involves renormalizable QHD models, which are 
characterized by a finite number of coupling constants and masses.
Since the number of parameters is finite, these models
provide a self-contained, Lorentz covariant, causal framework for
extrapolating known nuclear information to nuclear matter under
extreme conditions of density, temperature, and flow velocity.
Nevertheless, renormalizable models make specific assumptions about the
form of the interactions and the role of hadronic degrees of freedom
in the dynamics of the quantum vacuum.
These assumptions must be tested by comparing detailed calculations with
experiment.
Several such calculations have been performed in recent years,
which have indicated that the constraint of renormalizability is too
restrictive \cite{SW97r,Vac97r}.
This has prevented
the development of a systematic expansion procedure at finite density
and has also stimulated the search for alternatives to renormalizable models.

One guiding principle in the search for alternatives is that hadronic
theories should respect the underlying symmetries of QCD, such as chiral
symmetry.
Much work in recent years has focused on reconciling the successful
picture of nuclear structure mentioned earlier with chiral symmetry.
What has been learned is that the large scalar field present in nuclei,
which has its origin in the strong mid-range NN attraction,
is primarily a consequence of correlated two-pion exchange between the
nucleons \cite{DURSOr,LIN89r,LIN90r}.
This dynamics is implemented most efficiently by using a nonlinear
realization of the chiral symmetry to describe the pion--nucleon 
interactions\cite{CALLAN69a} and
by adding an additional, effective, scalar-isoscalar, 
chiral scalar field to simulate
the correlated two-pion exchange \cite{FURNSTAHL93,FST97r}.
One conclusion from this work is that even within the
context of hadronic field theories, one is led to the introduction of
{\em effective fields\/} to describe the relevant dynamics at 
large distances.
This observation, together with the difficulties encountered in the
consistent application of renormalizable models, suggests that 
{\it one should formulate
hadronic many-body theories as effective field theories from the outset}.

The scalar and vector meson fields will therefore be considered as effective
fields, and their dynamics can be implemented in two different ways: first,
by including these fields in an effective lagrangian, one has both an
obvious mean-field (factorized) limit, 
leading to a Dirac--Hartree description
similar to that discussed above, as well as the possibility of developing
systematic improvements to this approximation; alternatively, by including
these fields in an energy functional, one can apply many of the well-known
techniques of (nonrelativistic) density functional theory to a covariant
energy functional.

The application of effective relativistic field theory
to the nuclear many-body problem is in its adolescence, with new 
developments occurring rapidly.
Progress in this area will require the synthesis of ideas from
both many-body theory and relativistic quantum field theory.
The following sections describe in detail some of the recent progress and
likely future directions, such as the
application of these effective theories at the one-loop level
to electroweak interactions with nuclei;
the development of systematic techniques for going beyond one-loop to
include exchange, correlation, and vacuum effects; and studies of how
to extend the formalism to higher densities (and temperatures).
The role of chiral symmetry and the resulting pion
dynamics is also important, 
but we emphasize that the nuclear many-body problem involves aspects
that go beyond chiral symmetry, such as vector-meson dominance and the
broken scale invariance of QCD, which manifests itself in the dynamics
of the scalar-isoscalar sector.

\section{An Effective Chiral Lagrangian for Nuclei and Nuclear Matter}
\label{sec:tang}
\medskip

As discussed in Ref.~[\ref{FST97r}], we have recently constructed an
effective hadronic lagrangian consistent with the symmetries of 
quantum chromodynamics and intended for applications to finite-density 
nuclear systems \cite{TANGONEp,VMDp}.
The degrees of freedom are (valence) nucleons, pions, and the low-lying
non-Goldstone bosons, which account for the intermediate-range
nucleon--nucleon interactions and conveniently describe the nonvanishing
expectation values of bilinear products of nucleon fields.
Chiral symmetry is realized nonlinearly \cite{CALLAN69a},
with a light scalar meson
included as a chiral singlet to describe the mid-range
nucleon--nucleon attraction.
The low-energy electromagnetic structure of the nucleon is
described using vector-meson dominance, so
that \emph{ad hoc} form factors are not needed.

The effective lagrangian is expanded in powers of the fields and
their derivatives, with the terms organized using Georgi's
``naive dimensional analysis'' \cite{GEORGI93s}.
Results for finite nuclei and nuclear matter have been calculated
at one-baryon-loop order, using the single-nucleon structure
determined within the theory.
This leads essentially to a factorized parametrization of the 
ground-state energy functional.
Although the form of the scalar effective potential is motivated from
broken scale invariance \cite{TANGONEp},
we find that only the first few terms in an
expansion of powers of the scalar field are important at normal nuclear
densities, and thus the primary source of constraints on the scalar 
dynamics is the nuclear structure physics, which determines the
free parameters in the scalar potential.
These parameters implicitly contain the effects of many-body forces,
hadron substructure, and vacuum dynamics.
The parameters are obtained from fits to nuclear properties 
and show that naive 
dimensional analysis is a useful principle, and that a truncation
of the effective lagrangian at the first few powers of the fields
and their derivatives is justified \cite{FST97r,FSp00r}.
This is because the relevant expansion parameters are the ratios
of the meson mean fields in nuclei to the nucleon mass, and these
ratios are roughly 1/4 to 1/3 at the nuclear densities of interest.
Recently, we have verified that bulk and single-particle nuclear data
can determine roughly five, isoscalar, non-gradient parameters, one
gradient parameter, and one isovector parameter \cite{FSp00r}.

\section{Electroweak Interactions with Nuclei}
\label{sec:EWK}
\medskip

One of the advantages of the effective field theory is that the structure
of the hadrons can be included with increasing detail by adding more
and more nonrenormalizable interactions in a derivative expansion.
For example, the long-range parts of the single-nucleon electromagnetic
form factors (that is, magnetic moments and mean-square radii) can be
described using an effective lagrangian that includes vector-meson dominance
and direct couplings of photons to nucleons \cite{VMDp}.
The parameters describing the nucleon structure are adjusted to reproduce
free nucleon data (so that external form factors need not be introduced),
and the lagrangian dynamics then determines how this structure is 
modified inside the nucleus.
Moreover, with the DFT approach, accurate nuclear ground-state 
densities and
single-particle wave functions (for states near the Fermi surface) can be
obtained, and the electromagnetic current is automatically conserved by
virtue of the U(1) gauge invariance of the effective lagrangian
(and the resulting field equations).

Thus it is possible within this framework to study (e,e$'$N) reactions with
the same single-particle potentials in the initial and final state (thereby
ensuring current conservation), with realistic nuclear wave functions, and
with the medium-modified nucleon structure specified unambiguously.
While this approach may be too simple, it has the virtue of including all
of these effects simultaneously within the framework of a single lagrangian.
Moreover, meson-exchange currents consistent with gauge invariance can be
derived systematically, and one can extend the analysis to include coupled
channels as a way of introducing absorptive effects.
One of the challenges in this program is to extend the description of the
nucleon structure so that it is useful at the higher momentum transfers 
that are relevant to experiments at Jefferson Laboratory.

Initial studies of the weak currents within this framework focused
on axial-vector currents, including meson-exchange currents \cite{AXC}.
Although the leading exchange currents (in inverse powers of the nucleon mass)
can be deduced from $\mathrm{SU(2)}_L \times \mathrm{SU(2)}_R$ current algebra,
it is necessary to have a lagrangian with a
consistent power counting to extend these results to higher orders.
It is also known that in models based on a linear representation of the
chiral symmetry (such as the $\sigma$ model), 
it is impossible to satisfy the three constraints of PCAC,
the Goldberger--Treiman relation, and the correct charge algebra simultaneously
(at least at any tractable level of approximation) \cite{SERGEIp}.
For the nonlinear chiral lagrangian discussed above, however, PCAC is built
into the lagrangian, and the axial coupling between pions and nucleons
contains a free parameter, which allows all three constraints to be satisfied.
Moreover, as previously mentioned, the other parameters in the lagrangian can
be chosen to yield accurate nuclear wave functions that can be used to
compute matrix elements of these exchange currents within a single framework.
Finally, the pion--pion and pion--nucleon part of the lagrangian is exactly
the same as that used in chiral perturbation theory \cite{BIRA99}.

The weak vector and axial-vector currents have already been determined through
all orders of the pion field and up to two derivatives of pion fields,
together with contributions involving both rho mesons and pions.
(For computation of axial exchange currents, one requires only purely mesonic
terms and terms that are bilinear in the baryon fields.)
The corresponding weak charges reproduce the standard
charge algebra, and the Goldberger--Treiman relation is satisfied at
the tree level.
The two-body, axial-vector, meson-exchange amplitudes satisfy PCAC \cite{AXC}.
The remaining task is to deduce the exchange currents in forms that will
be useful in calculations using either relativistic (four-component) or
nonrelativistic (two-component) nuclear wave functions.
One can then analyze the exchange-current contributions in some selected
weak reactions and compare the results to earlier calculations.

\section{Calculations Beyond One-Loop Order}
\label{sec:EFTloop}
\medskip

One can also improve the description of the many-body dynamics within
the effective lagrangian approach.
At the one-loop level, the nucleons move in classical meson fields, and
the vacuum contributions are easy to handle because they can be separated
from the contributions of valence nucleons and then parametrized by 
polynomials in the meson fields in the lagrangian (or energy functional).
Although the one-loop calculations provide a realistic description of bulk
and single-particle nuclear properties, which can be understood within the
framework of density functional theory (see below),
it is still imperative to study exchange and correlation corrections
in a systematic way in order to make reliable predictions for other nuclear
observables.
Moreover, it is important to verify that the ``natural'' values of the
parameters obtained from fits at the one-loop level do not become unnatural
when higher-order, many-body effects are included.
Whereas the valence-nucleon contributions to these corrections are analogous
in structure to those in nonrelativistic many-body theory,
the vacuum dynamics is more complicated.
This is because one must now consider counterterms involving the nucleon
mass and wave function, as well as the nucleon--meson vertices, and one
must deal with ``mixed'' vacuum and valence terms, such as the strong
nuclear Lamb shift \cite{TWOLOOP}.

Recent work\cite{HU00} has focused 
on applying the loop expansion to the chiral lagrangian discussed above.
While we do not expect the loop expansion to be practical for many-body
calculations, it nevertheless provides a systematic framework for studying
the interplay of short-range (vacuum) and long-range (many-body) effects.
Moreover, the insight gained from the loop expansion has allowed us to extend
the analysis to more relevant approximations, such as those involving the
summation of rings and ladders.
What one finds (perhaps not surprisingly) is that in a large class of 
approximations, it is possible to separate the short-range contributions from
the long-range contributions, and to show that the former can be written in
terms of products of fields that are already present in the lagrangian.
In short, this implies that once one decides on a ``canonical'' form for the
lagrangian (which means that one chooses a set of field variables that 
eliminates redundant terms in a well-defined---but nonunique---fashion),
the resulting parameters implicitly include short-range effects {\it to all
orders\/} in the interaction, and these effects need not be calculated
explicitly.
The long-range, many-body effects from valence nucleons, which must be
calculated explicitly, resemble well-known contributions from nonrelativistic
many-body theory, like summations of ladders and rings.
The only differences are that the baryons now have Dirac wave functions, the
meson propagation is retarded, and there are modifications to the meson
propagation coming from nonlinear meson interactions that incorporate 
short-range physics.
Thus the nuclear dynamics can be organized in a fashion very similar to
the nonrelativistic nuclear many-body problem, with the only differences
being the relativistic modifications mentioned above and a small number
of unknown parameters that describe the short-range dynamics (which is
manifested through many-body forces).
This new approach to the separation of ``valence'' and ``vacuum'' effects
solves a problem that has existed in QHD for more than 20 years.

Obviously, there is still a tremendous amount of work that must be done
within this program.
Our preliminary calculations at the two-loop level and in a simplified
Dirac--Brueckner--Hartree--Fock calculation show that it remains possible
to reproduce the empirical properties of nuclear matter with parameters that
are natural; this is further evidence of the dominance of the Hartree
contributions to these parameters.
Nevertheless, these Brueckner calculations must be improved, as there are
still several ambiguities that must be removed: a consistent
cutoff procedure, which truncates integrals systematically at the scale
of the ``heavy'' non-Goldstone-boson masses must be implemented (and
checked for sensitivity), and the retardation and short-range modifications
to the meson propagators must be included.
A systematic method for constructing a canonical form of the lagrangian
(with no redundant terms) must also be found.
One must also develop the formulation of ``conserving 
approximations''\cite{BAYM61,BAYM62} 
that go beyond the simple one-loop level and that allow
for the inclusion of retarded interactions.

\section{Energy Functional Analysis of Mean-Field Theories of Nuclei}
\label{sec:bodmer}
\medskip

As an alternative approach to the relativistic nuclear many-body problem,
we consider an energy functional that depends on valence-nucleon wave
functions and classical scalar and vector fields.
Extremization of this functional leads to coupled equations for finite
nuclei and nuclear matter, and the success of relativistic mean-field
models discussed earlier shows that these variables allow an efficient 
description of bulk and single-particle nuclear properties.

Although the energy functional contains classical meson fields,
this framework can accommodate physics beyond the simple Hartree 
(or one-baryon-loop) approximation.
This is achieved by combining aspects of both 
density functional theory\cite{DREIZLER90p,SPEICHER92p,SCHMID95,ENGEL95}
(DFT) 
and effective field theory\cite{WEINBERG67,GEORGI93u,WEINBERG95p} (EFT).
In a DFT formulation of the relativistic nuclear 
many-body problem, the central object is an energy functional of scalar and 
vector densities (or more generally, vector four-currents).
Extremization of the functional gives rise to Dirac equations 
for occupied orbitals with {\it local\/} scalar and vector potentials, 
not only in the Hartree approximation, but in the general case as well.
(Note that the Dirac eigenvalues do not correspond precisely
to physical energies in the general case \cite{DREIZLER90p}).
Rather than work solely with the densities,
we can introduce auxiliary variables corresponding to the local potentials,
so that the energy functional depends also on meson fields.
The resulting DFT formulation takes the form of a Hartree calculation, but 
correlation effects can be included, {\it if\/} the proper density 
functional can be found.
Our procedure is analogous to the well-known Kohn--Sham\cite{KOHN65}
approach in DFT, with the local meson fields playing the role of Kohn--Sham
potentials; by introducing nonlinear couplings between these fields, we can
implicitly include density dependence in the single-particle potentials.

Moreover, by introducing the meson fields, we can incorporate the ideas 
of EFT.
The exact energy functional has kinetic energy and Hartree parts 
(which are combined in the relativistic formulation) plus 
an ``exchange-correlation'' functional, which is a nonlocal, nonanalytic 
functional of the densities that contains all the other many-body and 
relativistic effects.
At the present stage of our investigations, we
do not try to construct the latter functional explicitly from a 
lagrangian (which would be equivalent to solving the full many-body problem),
but instead approximate the functional using an expansion in classical meson
fields and their derivatives.
The parameters introduced in the expansion can be fit to experiment, and 
if we have a systematic way to truncate the expansion, the framework is 
predictive.
Thus a conventional mean-field energy functional fit directly to nuclear 
properties, if allowed to be sufficiently general, will automatically 
incorporate effects beyond the Hartree approximation, such as those due to
short-range correlations.
These observations serve to justify existing relativistic mean-field models
containing nonlinear meson self-interactions, which are successful, but 
which involve lagrangians or energy functionals that have customarily
been truncated at some low order without any justification.

We rely on the special characteristics of nuclear ground states
in a relativistic formulation, namely, that the mean 
scalar and vector potentials $\Phi$ and $W$ are large on nuclear energy 
scales but are small compared to the nucleon mass 
$M$ \cite{TANGONEp,BODMER91}.
This implies that the ratios $\Phi/M$ and $W/M$ provide useful expansion 
parameters.
Moreover, as is illustrated in Dirac--Brueckner--Hartree--Fock (DBHF)
calculations \cite{HOROWITZ87p,TERHAAR87,MACHLEIDT89},
the scalar and vector potentials (or self-energies) are nearly 
state independent and are dominated by the Hartree contributions.
Thus the Hartree contributions to the energy functional should dominate,
and an expansion of the exchange-correlation functional in terms of mean
fields should be a reasonable approximation.
This ``Hartree dominance'' also implies that it should be a good 
approximation to associate the single-particle Dirac eigenvalues with the 
observed nuclear energy levels, at least for states near the Fermi 
surface \cite{DREIZLER90p}.
Of course, the mean-field expansion cannot accommodate all of the nonlocal 
and nonanalytic aspects of the exchange-correlation functional, and 
important future work must be done to
see how these additional effects can be best incorporated.

Given a suitable truncation scheme, the first step is to analyze a general
energy functional (through some level of truncation) to determine the
characteristics that generate successful nuclear phenomenology.
One wants to accurately reproduce nuclear charge densities, binding-energy
systematics, and single-particle energy levels.
If one recalls that the Kohn--Sham approach is formulated to reproduce
precisely the ground-state density, and that the relativistic Hartree
contributions are expected to dominate the Dirac single-particle potentials,
these observables are precisely the ones for which meaningful comparisons
with experiment should be possible.
Moreover, experience has shown that these observables can be replaced by
a set of nuclear matter properties plus constraints on the meson masses.
Finally, it is possible to analytically invert the field equations to
solve directly for the model parameters in terms of the nuclear matter
input properties \cite{BODMERp}.
This allows for a systematic and complete study of the parameter space, so
that parameter sets that accurately reproduce nuclear observables can be
found, and models that fail to reproduce nuclear properties can be excluded.
For example, one learns that favored parameter sets typically involve small
but significant nonlinear meson self-interactions.
What remains to be done is to identify optimal linear combinations of
parameters that are the most tightly constrained by the data and use these
to produce some ``favored'' parameter sets.

One of the primary directions for future study will involve the assumption
of ``naturalness,'' namely, that once the appropriate mass scales have been
identified, the coefficients of various terms in
the energy functional, when expressed in dimensionless form,
should all be of order unity.
Naturalness allows us to estimate the size of the terms omitted from the
energy functional and also implies that one should include all possible
terms allowed by the symmetries through a given level of truncation;  thus
nonlinear vector--vector and vector--scalar interactions should be as
important as scalar--scalar interactions in producing accurate nuclear
observables, and we indeed find that this is the case.
Nevertheless, because the general energy functional contains all powers
in the fields and densities, one has great freedom to make field
redefinitions, and the relevant question is which representation of the
interactions leads to the most natural and efficient truncation scheme.
We have found that simple nucleon--meson Yukawa couplings supplemented by
meson self-interactions is one way to provide a description with natural
parameters, but it may also be possible to use more complicated
meson--nucleon couplings (like ${\overline\psi}\psi\phi^n$) or higher 
powers of the nuclear scalar and vector densities 
(``contact terms'') \cite{PC97}.

We also plan to investigate the relationship between the energy functional
and underlying effective lagrangians (such as the one discussed in
Section \ref{sec:tang}) by performing microscopic calculations beyond the
Hartree level.
Although there are some outstanding formal issues in the derivation of
relativistic density functional theory (such as the existence of a
relativistic Hohenberg--Kohn theorem, or variational 
principle\cite{SPEICHER92p}), the primary technical issue is how
best to approximate the exchange, correlation, and vacuum corrections in
terms of the Kohn--Sham potentials.
For example, exchange terms introduce momentum dependence into the baryon
self-energies in infinite matter, but the Kohn--Sham potentials used to 
approximate these self-energies should be state independent.
Is there an optimal momentum to use to approximate the state dependence?
Is it better to introduce small nonlocalities?
Preliminary studies on these questions have been done, but these issues are
still a long way from being resolved.

\section{The Nuclear Symmetry Energy in Covariant Density Functionals}
\label{sec:isospin}
\medskip

A power-counting analysis shows that only one isovector parameter in covariant
mean-field density functionals is determined by conventional fits to nuclear
properties \cite{FSp00r}.
This implies that the density dependence of the symmetry energy is not
constrained by a good fit, but only some average value is.
The usual parametrization of isovector interactions in terms of simple 
rho-meson exchange leads to a much larger density 
dependence than is obtained in most nonrelativistic Skyrme models.
An observable consequence is a significantly larger prediction for the neutron
skin in heavy nuclei.
Relativistic and nonrelativistic Brueckner calculations of asymmetric nuclear
matter using realistic interactions all predict a density dependence consistent
with the lower values, which suggests that the covariant parametrization may
be deficient.
It is important
to examine the effect on the symmetry energy of more general isovector
terms and of explicit pion-exchange contributions to the energy functional.
These results could have impact on the structure of neutron stars.

\section*{Acknowledgements}
\smallskip

This work was supported in part by the US Department of Energy under contract
No.~DE-FG02-87ER40365.

%

\begin{thebibliography}{99}
%
\bibitem{SW86r} B. D. Serot and J. D. Walecka,
                  Adv.\ Nucl.\ Phys.\ {\bf 16}, 1 (1986),
                  and references therein.
%
\bibitem{BDS92r} B. D. Serot, Rep.\ Prog.\ Phys.\ {\bf 55}, 1855 (1992).
%
\bibitem{FST97r} \label{FST97r}
            R. J. Furnstahl, B. D. Serot, and H.-B. Tang, Nucl.\ Phys.\
            {\bf A615}, 441 (1997); {\bf A640}, 505 (1998) (E).
%
\bibitem{SW97r} B. D. Serot and J. D. Walecka, Int.\ J.\ Mod.\ Phys.\ E
            {\bf 6}, 515 (1997), and references therein.
%
\bibitem{FSp00r} R. J. Furnstahl and B. D. Serot, Nucl.\ Phys.\ {\bf A671},
             447 (2000).
%
\bibitem{FSL00r} R. J. Furnstahl and B. D. Serot, Nucl.\ Phys.\ {\bf A673},
             298 (2000).
%
\bibitem{EvRev00r} R. J. Furnstahl and B. D. Serot, Comm.\ Mod.\ Phys.\
             {\bf 2}, A23 (2000).
%
\bibitem{SW00r} {\sc Effective Field Theory in Nuclear Many-Body
             Physics}, B. D. Serot and J. D. Walecka,
             in: {\it 150 Years of Quantum Many-Body Theory\/}: A
             Festschrift in Honour of the 65${}^{\rm th}$
             Birthdays of John W. Clark,
             Alpo J. Kallio, Manfred L. Ristig, and Sergio Rosati;
             R. F. Bishop, K. A. Gernoth, and N. R. Walet, eds.~(World
             Scientific, Singapore, 2001), p.~203.
%
\bibitem{FRS98r} R. J. Furnstahl, J. J. Rusnak, and B. D. Serot,
             Nucl.\ Phys.\ {\bf A632}, 607 (1998).
%
\bibitem{COHEN91r} T. D. Cohen, R. J. Furnstahl, and D. K. Griegel, Phys.\ 
            Rev.\ Lett.\ {\bf 67}, 961 (1991).
%
\bibitem{FURNST92r} R. J. Furnstahl, D. K. Griegel, and T. D. Cohen,
           Phys.\ Rev.\ C {\bf 46}, 1507 (1992).
%
\bibitem{Vac97r} R. J. Furnstahl, B. D. Serot, H.-B. Tang, Nucl.\ Phys.\
             {\bf A618}, 446 (1997).
%
\bibitem{DURSOr}J. W. Durso, A. D. Jackson, and B.~J.\ VerWest,
          Nucl.\ Phys.\ {\bf A345}, 471 (1980).
%
\bibitem{LIN89r}W. Lin and B. D. Serot, Phys.\ Lett.\ {\bf 233B}, 23 (1989).  
%
\bibitem{LIN90r} W. Lin and B. D. Serot,
              Nucl.\ Phys.\ {\bf A512}, 637 (1990).
%
\bibitem{CALLAN69a} C. Callan, S. Coleman, J. Wess, and B. Zumino,
       Phys.\ Rev.\ {\bf 177}, 2247 (1969).
%
\bibitem{FURNSTAHL93} R. J. Furnstahl and B. D. Serot,
             Phys.\ Rev.\  C {\bf 47}, 2338 (1993);
             Phys.\ Lett.\ B {\bf 316}, 12 (1993).
%
\bibitem{TANGONEp} R. J. Furnstahl, H.-B. Tang, and B. D. Serot, Phys.\ 
       Rev.\ C {\bf 52}, 1368 (1995).
%
\bibitem{VMDp} R. J. Furnstahl, B. D. Serot, and H.-B. Tang, Nucl.\ Phys.\
       {\bf A615}, 441 (1997).
%
\bibitem{GEORGI93s} H. Georgi, Phys.\ Lett.\ {\bf 298B}, 187 (1993).
%
\bibitem{AXC} S. M. Ananyan, B. D. Serot, and J. D. Walecka
             (2001), submitted to Physical Review C.
%
\bibitem{SERGEIp}  S. M. Ananyan,  Phys.\ Rev.\ C {\bf 57}, 2669 (1998).
%
\bibitem{BIRA99} U. van Kolck, Prog.\ Part.\ Nucl.\ Phys.\ {\bf 43}
            (1999) 337.
%
\bibitem{TWOLOOP} R. J. Furnstahl, R. J. Perry, and B. D. Serot,
       Phys.\ Rev.\ C {\bf 40}, 321 (1989).
%
\bibitem{HU00} Y. Hu, Ph.D. Thesis, Indiana University (2000).
%
\bibitem{BAYM61} G. Baym and L. P. Kadanoff, Phys.\ Rev.\ {\bf 124},
       287 (1961).
%
\bibitem{BAYM62} G. Baym, Phys.\ Rev.\ {\bf 127}, 1391 (1962).
%
\bibitem{DREIZLER90p}R. M. Dreizler and E. K. U. Gross,
      {\it Density Functional Theory\/} (Springer, Berlin, 1990).
%
\bibitem{SPEICHER92p}C. Speicher, R. M. Dreizler, and E. Engel,
     Ann.\ Phys.\ (NY)  {\bf 213}, 312 (1992).
%
\bibitem{SCHMID95}R. N. Schmid, E. Engel, and R. M. Dreizler,
       Phys.\ Rev.\ C {\bf 52}, 164 (1995).
%
\bibitem{ENGEL95}E. Engel, H. M\"uller, C. Speicher, and R. M. Dreizler,
       in {\it Density Functional Theory}, NATO Advanced Science Institute
       Series B, vol.~337, E. K. U. Gross and R. M. Dreizler,
       eds.~(Plenum, New York, 1995).
%
\bibitem{WEINBERG67}S. Weinberg, Phys.\ Rev.\ Lett.\ {\bf 18}, 188 (1967).  
%
\bibitem{GEORGI93u}H. Georgi, Ann.\ Rev.\ Nucl.\ Part.\ Sci.\ {\bf 43},
       209 (1993).
%
\bibitem{WEINBERG95p}S. Weinberg, {\it The Quantum Theory of Fields, vol.~I:
       Foundations} (Cambridge Univ. Press, New York, 1995).
%
\bibitem{KOHN65} W. Kohn and L. J. Sham, Phys.\ Rev.\ A {\bf 140}, 1133
      (1965).
%
\bibitem{BODMER91}A. R. Bodmer, Nucl.\ Phys.\ {\bf A526}, 703 (1991).
%
\bibitem{HOROWITZ87p}C. J. Horowitz and B. D. Serot, Nucl.\ Phys.\ {\bf A464},
         613 (1987);  {\bf A473}, 760 (1987) (E).
%
\bibitem{TERHAAR87}B. ter Haar and R. Malfliet, Phys.\ Rep.\ {\bf 149},
         207 (1987).
%
\bibitem{MACHLEIDT89}R. Machleidt, Adv.\ Nucl.\ Phys.\ {\bf 19},
            189 (1989).
%
\bibitem{BODMERp} R. J. Furnstahl, B. D. Serot, and H.-B. Tang,
            Nucl.\ Phys.\ {\bf A598}, 539 (1996).
%
\bibitem{PC97} J. J. Rusnak and R. J. Furnstahl, Nucl.\ Phys.\
            {\bf A627}, 495 (1997).
%
\end{thebibliography}
\end{document}